# Electrically tunable second harmonic generation in atomically thin ReS$_2$

*Jing Wang,[†,‡] Nannan Han,[§,‡] Zheng-Dong Luo,[#] Mingwen Zhang,[†] Xiaoqing Chen,[†] Yan Liu,[#] Yue Hao,[#] Jianlin Zhao,[†] Xuetao Gan,*[,†]*

[†]Key Laboratory of Light Field Manipulation and Information Acquisition, Ministry of Industry and Information Technology, and Shaanxi Key Laboratory of Optical Information Technology, School of Physical Science and Technology, Northwestern Polytechnical University, Xi'an, 710129 China

[§]Frontiers Science Center for Flexible Electronics, Institute of Flexible Electronics, Northwestern Polytechnical University, Xi'an 710072, China

[#]Wide Bandgap Semiconductor Technology Disciplines State Key Laboratory, School of Microelectronics, Xidian University, Xi'an 710071, China

**ABSTRACT:** Electrical tuning of second-order nonlinearity in optical materials is attractive to strengthen and expand the functionalities of nonlinear optical technologies, though its implementation remains elusive. Here, we report the electrically tunable second-order nonlinearity in atomically thin $ReS_2$ flakes benefiting from their distorted 1T crystal structure and interlayer charge transfer. Enabled by the efficient electrostatic control of the few-atomic-layer $ReS_2$, we show that second harmonic generation (SHG) can be induced in odd-number-layered $ReS_2$ flakes which are centrosymmetric and thus without intrinsic SHG. Moreover, the SHG can be precisely modulated by the electric field, reversibly switching from almost zero to an amplitude more than one order of magnitude stronger than that of the monolayer $MoS_2$. For the even-number-layered $ReS_2$ flakes with the intrinsic SHG, the external electric field could be leveraged to enhance the SHG. We further perform the first-principles calculations which suggest that the modification of in-plane second-order hyperpolarizability by the redistributed interlayer-transferring charges in the distorted 1T crystal structure underlies the electrically tunable SHG in $ReS_2$. With its active SHG tunability while using the facile electrostatic control, our work may further expand the nonlinear optoelectronic functions of two-dimensional materials for developing electrically controllable nonlinear optoelectronic devices.

**KEYWORDS:** rhenium disulfide, second harmonic generation, first-principles calculations, electrical tuning, interlayer charge transfer, asymmetric charge density distribution

Second-order nonlinearity in optical materials is fundamental for a large variety of modern optical applications, including but not limited to exploiting lasers in awkward wavelengths,[1,2] manufacturing Pockels electro-optic modulators,[3] generating entangled photons,[4] developing advanced spectroscopy for surface/interface and crystalline physics.[5] A feasibly tunable second-order nonlinearity under external modulations would further augment and/or improve the functionalities of these devices. An ultimate solution to this end would be a reconfigurable functional material whose second-order nonlinearity can be customized on demand, so that a high degree of functionalities can be realized in related nonlinear optical devices. Unfortunately, to the best of our knowledge, an efficient approach achieving the tunable second-order nonlinearity remains elusive.

Recently, two-dimensional (2D) materials have been demonstrated with great electric-tunable optical and electronic properties benefiting from their thicknesses of few-atomic layers.[6-8] For example, graphene's optical absorption could be prohibited by electrostatic control.[9] The polarity of $WSe_2$ layers has been shown to be dynamically controlled for constructing reconfigurable logic circuits.[10] These imply that the 2D materials could be a promising candidate for realizing tunable second-order optical nonlinearity. There are some reported trials.[11-14] By electrically switching the exciton and trion oscillators in the monolayer $WSe_2$, second harmonic generation (SHG) on-resonance with the exciton was modulated.[12] Though it only happens on the narrowband exciton peak, such effect is prohibited in bulk materials due to the absence of robust exciton states. In a centrosymmetric bilayer transition metal dichalcogenide (TMD), which has no intrinsic second-order nonlinearity, SHGs could be induced by external modulations, such as the actuation of an external electric field or unbalanced charge accumulations.[13,14] Unfortunately, such induced SHGs are typically several orders of magnitude weaker than those from their monolayer counterparts, which hinders their possible applications for realistic nonlinear optical devices.

In this work, we report electrically tunable SHG in atomically thin ReS$_2$, which is valid over a broad spectral range. ReS$_2$, a group VII-TMDs, has a distorted 1T (1T') lattice structure, as demonstrated clearly in previous reports by the transmission electron microscope technique.[15,16] Governed by the lattice structure, it presents distinct optical and electronic properties in comparison with group VI-TMDs (*e.g.*, MoS$_2$, WS$_2$).[17-20] As reported in our previous work, ReS$_2$ with specific layer-numbers exhibits extraordinary SHG properties.[21] To be specific, no SHG response is observed in monolayer ReS$_2$ while bilayer ReS$_2$ shows strong SHG, which is opposite to the layer dependence of SHG of layered group VI-TMDs.[22-24] This indicates the interlayer coupling between ReS$_2$ monolayers with a distorted 1T stacking order is essential to generate the second-order nonlinear polarizability for SHG, though it is absent in an individual ReS$_2$ monolayer. Considering that the interlayer coupling is governed by the relatively weak van der Waals interaction, ReS$_2$ could be exploited to realize tunable SHG by perturbing the interlayer coupling. In this work, we show that from ReS$_2$ flakes with odd layer-numbers, such as trilayer and five-layer, which intrinsically have no SHG, strong SHG response emerge with the actuation of an electric field. The achieved SHG strength is more than one order of magnitude stronger than that of the monolayer MoS$_2$, which is well recognized as an optical material with strong second-order nonlinearity. In the even-number-layered ReS$_2$ flakes with intrinsic SHG, an external electric field could be used to modulate and even enhance the SHG response by several times. The first-principles calculations are carried out to explain the electric-tunable SHG in ReS$_2$ layers, which suggests that it can be attributed to the electric field-induced in-plane asymmetric charge distribution governed by the distorted 1T lattice structure. The strong and tunable second-order nonlinearity in atomically thin ReS$_2$ flakes could lead to electrically controlled nonlinear optoelectronic devices.

**RESULTS AND DISCUSSION**

To carry out the electrically tuned SHG in atomically thin ReS$_2$ flakes, we fabricated ReS$_2$ field effect transistors (FETs) with different channel thicknesses. Figure 1a shows the structure diagram of a trilayer ReS$_2$ FET. In the device fabrication, few-layer ReS$_2$ flakes were mechanically exfoliated from the bulk ReS$_2$ single crystal using a polydimethylsiloxane (PDMS) stamp, which were then transferred onto the heavily doped Si substrate covered with a 300 nm thick SiO$_2$ layer. Here, the layer numbers of the ReS$_2$ flakes are measured using the ultralow frequency Raman spectroscopy (see the Supporting Information) and the atomic force microscopy (AFM). Two Au electrodes with a thickness of 100 nm were then transferred on the ReS$_2$ flake acting as the drain and source electrodes.[25-29] The heavily doped Si substrate could function as a bottom electrical gate. By applying an electrical voltage between the heavily doped Si substrate and the source electrode, *i.e.* gate voltage, an electric field could be vertically applied across the ReS$_2$ plane, which is expected to modify the electronic and optical properties of the ReS$_2$ flake. All the electrical and optical measurements were implemented under an inert nitrogen environment at room temperature, which could effectively eliminate the influence of molecule absorption in the ambient, such as water, oxygen molecules and other impurities.

Figure 1b shows the optical microscope image of the fabricated trilayer ReS$_2$ FET. To characterize its electrical performances, we apply source-drain voltages ($V_{DS}$) between -0.05 V and 0.05 V and measure the source-drain currents ($I_{DS}$) under different gate voltages ($V_G$). The results are shown in Figure 1c. The linear dependence between $I_{DS}$ and $V_{DS}$ at different $V_G$ indicates the good Ohmic contacts between the electrodes and the ReS$_2$ channel. With $V_{DS}$ fixed at 0.1 V, the $I_{DS}$ as a function of the $V_G$ (scanned from -60 V to 60 V) is acquired, as shown in Figure 1d. The channel gradually switched from the off state to the on state starting from the threshold voltage $V_T \approx$ -20 V with the a large on/off ratio of $10^3$, which indicates a typical n-type semiconducting behavior of the trilayer ReS$_2$ and the excellent electrostatic control.

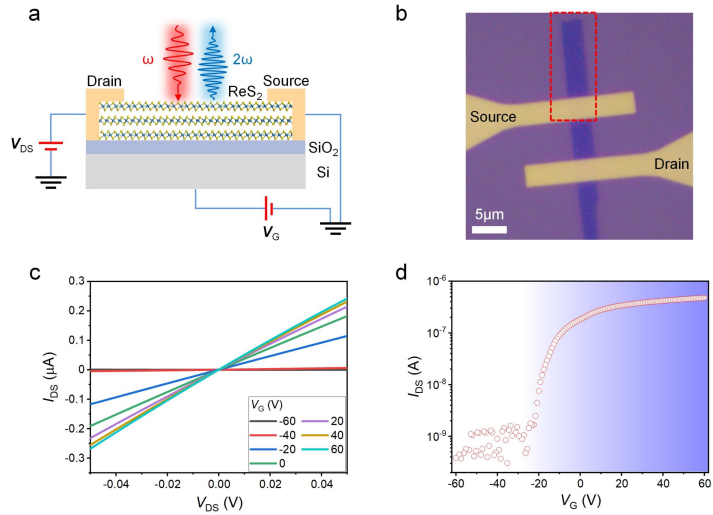

**Figure 1.** (a) Structure diagram of the trilayer ReS$_2$ FET. Excitation at $\omega$ (red arrow) generates second-harmonic radiation at $2\omega$ (blue arrow). (b) Optical microscope image of the fabricated trilayer ReS$_2$ FET. (c) Output characteristics ($I_{DS}$-$V_{DS}$) of the trilayer ReS$_2$ FET under different gate voltages $V_G$, showing good Ohmic contacts between the electrodes and the ReS$_2$ channel. (d) Log scale plot of the transfer curve ($I_{DS}$-$V_G$) of the trilayer ReS$_2$ FET with $V_{DS}$ = 0.1 V.

The SHG of the trilayer ReS$_2$ is evaluated using a home-built vertical microscope setup with the reflection geometry. A picosecond pulsed laser from an optical parametric oscillator (OPO) is employed as the fundamental pump radiation, whose wavelength could be tuned to examine the wavelength dependence of the SHG. The pump laser is first fixed at the wavelength of 1500 nm, which is focused by a 50× objective lens with a numerical aperture of 0.75 into a spot size about 2 μm. SHG response from the ReS$_2$ could be collected by the same objective lens, which is then examined by a spectrometer mounted with a cooled silicon CCD. The collected spectra from the ReS$_2$ at the range around the half of the pump wavelength are plotted in Figure 2a.

To detect the intrinsic SHG response of the ReS$_2$ layers, SHG response of the original trilayer ReS$_2$ flake (exfoliated on the PDMS) is examined first. As shown in Figure 2a, there is no measurable SHG signal, indicating that the intrinsic SHG of the trilayer ReS$_2$ is determined by its centrosymmetric crystal structure.[21] Next, SHG response of the fabricated trilayer ReS$_2$ FET is measured. Upon the application of $V_G$ = -60 V, a strong SHG signal clearly emerges from the ReS$_2$ channel between the two electrodes, see Figure 2a. Here, both the source and drain electrodes remain grounded during the SHG measurements to eliminate the possibility of current-induced SHG.[30,31] To verify the SHG characteristics of the detected radiation, we measure the SHG dependence on the pump power under $V_G$ = -60 V, as shown in the log-log plot of Figure 2b. A nearly quadratic dependence on the pump power with a slope of 1.9 ± 0.1 is obtained, implying the second-order nonlinear optical process. The laser polarization dependence of the parallel component of SHG intensity is also measured under $V_G$ = -60 V, which shows a twisted dumbbell shape (see the inset in Figure 2b). This result is same as that obtained from a ReS$_2$ flake with the even layer-number, as reported in our previous work,[21] which is determined by the distorted 1T lattice structure. The above results illustrate that an external electric field normal to the plane of the ReS$_2$ flake could effectively induce second-order nonlinearity in spite of the intrinsic centrosymmetry.

To estimate the strength of the induced second-order nonlinearity from the trilayer ReS$_2$, we also measure the SHG from a monolayer MoS$_2$ under the same experiment condition. The result is shown as the blue curve in Figure 2a. Under the same condition, the SHG intensity induced under $V_G$ = -60 V in trilayer $ReS_2$ is far greater than the intrinsic SHG intensity of the monolayer $MoS_2$, presenting a factor of 15 times. Considering that the monolayer $MoS_2$ is well recognized as one of the 2D materials with significantly strong second-order nonlinearity,[22,23,32-35] it is concluded that the actuation of electric field could yield second-order nonlinearity in trilayer $ReS_2$. On the contrary, in the reported electric field induced SHG in bilayer $MoS_2$ and $WSe_2$, the strengths are several orders of magnitude smaller than those from their monolayer counterparts.[13,14]

To explore the physical origin of the electric field enabled SHG from the trilayer $ReS_2$, we plot the measured SHG signals as a function of the gate voltages. Limited by the thin $SiO_2$ dielectric layer, significant leakage currents and brokedown of the dielectric layer are observed when the gate voltage exceeds ±60 V. This brokedown damage is irreversible and affects the subsequent measurement of the devices. Therefore, in our experiments, the maximum gate voltage is only applied as ±60 V. When $V_G$ > 0 V, the intensities of SHG increase slightly first and then decrease slightly. While when $V_G$ < 0 V, the intensities of SHG decrease to zero first and then increase dramatically. Note that the zero-SHG is not obtained at $V_G$ = 0 V, instead at $V_G$ ≈ -20 V. This $V_G$ dependence will be explained in the following text.

The pump wavelength dependence of the electrically enabled SHG in trilayer $ReS_2$ is also measured. The SHG intensities are monitored when the wavelength of the OPO pulsed laser is changed from 1500 nm to 1600 nm with a step of 5 nm. For each pump wavelength, the SHG signals are measured with $V_G$ scanning from -60 V to 60 V, as shown in Figure 2d. For different pump wavelengths, the electric field enabled SHGs demonstrate the similar function of the gate voltages. They all present rapidly increased values at the negative gate voltages. It reveals that the SHG in trilayer ReS$_2$ can be tuned electrically over a broad spectral range, which is different from the electrically controlled SHG enabled by tuned exciton behavior in monolayer WSe$_2$.[12] This experimental result not only indicates that the electrically tunable SHG in atomically thin ReS$_2$ is not originated from the exciton resonance, but also promises great potentials in strengthening and expanding the functionalities of nonlinear optical technologies using atomically thin ReS$_2$.

Finally, under different gate voltages, we carry out the SHG spatial mappings (pumped at 1500 nm) over the ReS$_2$ region marked in the red dashed box in Figure 1b, as shown in Figure 2e. The global Si back gate provides a uniform electric field over the whole ReS$_2$ flake, which results in the homogeneous SHG distribution under each gate voltages. Their $V_G$-dependences are consistent with the result shown in Figure 2c, which show the largest value at $V_G$ = -60 V and zero-SHG at $V_G$ = -20 V. It further indicates the reliable controllability of second-order nonlinearity in trilayer ReS$_2$ by an external electric field.

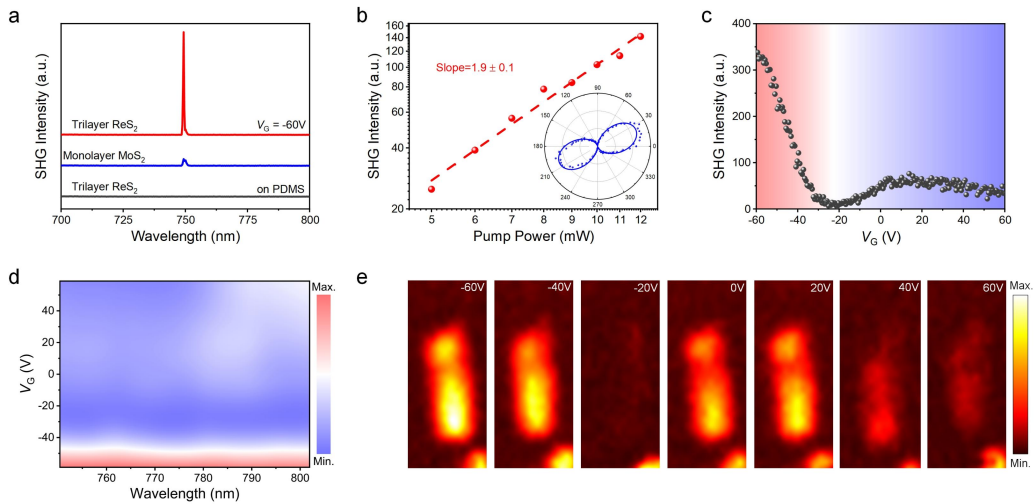

**Figure 2.** (a) SHG spectra measured from the trilayer ReS$_2$ on the PDMS substrate (black), the trilayer ReS$_2$ FET with $V_G$ = -60 V (red), and the monolayer MoS$_2$ (blue) under the same

experiment conditions. The spectra are shifted vertically for clarity and normalized by the data acquired from monolayer MoS$_2$. (b) Log-log plot of the pump power dependence of SHG from the trilayer ReS$_2$ FET at $V_G$ = -60 V. The inset is the polar plot of parallel components of SHG intensities with respect to the polarization of the pump laser. (c) Gate voltage dependence of the SHG intensities in the trilayer ReS$_2$ FET. (d) SHG intensities as a function of the gate voltage and the pump wavelength in the trilayer ReS$_2$ FET. (e) SHG spatial mapping images of the trilayer ReS$_2$ region marked by the red dashed box in Figure 1b under different gate voltages.

To explicate the above experiment results, the first-principles calculations are utilized. When the negative gate voltage is applied, *i.e.* $V_G$ < 0 V, electrons would move in the opposite direction to the electric field. In trilayer ReS$_2$, due to the charge transfer among the three monolayers stacked with a distorted 1T structure order, the charge density will be redistributed, as shown in Figure 3a, where red (green) region means gaining (losing) electrons. On the whole, the charge distribution will change from the state of central inversion symmetry to the state of inversion symmetry broken across the three monolayers under the negative gate voltage, and the asymmetry increases linearly with the increased magnitude of the negative gate voltage. With the vertical electric field across the multi-layer ReS$_2$, charges transfer across the multiple ReS$_2$ layers vertically, which gives rise to an asymmetry along the vertical direction. Benefiting from the distorted 1T lattice structure of ReS$_2$, the vertical asymmetric charge distribution could yield the in-plane component of the second-order nonlinear hyperpolarizability. When the electrically gated multi-layer ReS$_2$ is pumped by an optical field polarized along the in-plane direction, considerably tunable SHG could be realized. By contrast, for the group VI-TMDs (*e.g.*, MoS$_2$, WSe$_2$) without the distorted lattice structure, even there are considerable vertical electric fields applied across them,[13,14] the electrically tuned SHG is much weaker than the values obtained in the ReS$_2$ results. Hence, the tunable SHG in ReS$_2$ could attributed to its distorted lattice structure. As a result, with the broken inversion symmetry, strong second-order polarizability is yielded and increases linearly with the enlarged magnitude of the negative gate voltage. It is consistent with the experiment results shown in Figure 2c that the ReS$_2$ flake under a downward electric field presents strong SHG signals. When the negative $V_G$ is decreased to -60 V gradually, the obtained SHG intensities increase with a linear function.

Note that, for the results shown in Figure 2c, the zero-SHG takes place under $V_G \approx$ -20 V, which is not at $V_G$ = 0 V. This could be attributed to the slight n-type doping of ReS$_2$ by the SiO$_2$ substrate at $V_G$ = 0 V,[36,37] which induces an interface layer with negative charges in the bottom ReS$_2$ monolayer. Hence, in the trilayer ReS$_2$, the unbalanced charge distributions over the first and third layers could break the inversion symmetry and generate a second-order polarization with the illuminated optical field, which yields a SHG signal. When the negative $V_G$ is applied from 0 V to -20 V gradually, the downward external electric field would drift the substrate-doped negative charges in the bottom ReS$_2$ layer to the top layers, which recovers the inversion symmetry of the trilayer ReS$_2$ gradually. Hence, the SHG signal obtained at $V_G$ = 0 V diminishes to zero when the $V_G$ is close to -20 V. It agrees well with the threshold gate voltage in the transfer characteristic, which shows the depletion of doped electrons at $V_G \approx$ -20 V. If the $V_G$ is then applied from -20 V to -60 V, the external electric field would redistribute the charges across the three ReS$_2$ layers, resulting in net positive (negative) charges on the bottom (top) ReS$_2$ layers. It therefore breaks the inversion symmetry again. Consequently, the SHG emerges and increases significantly due to the strengthened interlayer charge transfer.

When the positive gate voltage is applied, *i.e.* $V_G$ > 0 V, the yielded external electric field points to the upward direction. It would strengthen the accumulation of negative charges on the bottom ReS$_2$ layer, and enhance the SHG signal enabled by the substrate-doping effect. As shown in Figure 2c, the SHG intensities increase slightly at $V_G$ around 15 V compared with that obtained at $V_G$ = 0 V. However, the SHG intensities decrease slightly under $V_G$ from 20 V to 60 V. Part of the reason is that electron accumulation layers will form on both Re and S atom layers as the conduction band minimum is contributed by both of them according to our calculated band structures shown in Figure 3b. Because the densities of the band edge states are so high that practically the entire space charges are localized almost in the bottom layer. The electrons accumulated on S atoms in the bottom layer will act as the charge shielding layer that prevents the further penetration of the electric field beyond this accumulation plane, as schematically shown in Figure 3c. Macroscopically, the bond charge distribution in the trilayer ReS$_2$ will hardly further affected by the external electric field though the gate voltage is increased continuously. Therefore, the SHG intensities should be almost unchanged under the positive gate voltages. However, slight decrease of SHG is observed when $V_G$ is increased from 20 V to larger values. The specific physical mechanism of this decreased SHG is still not clear yet. In future studies, it is interesting to apply even higher electric field to determine the further variation trend of the SHG signal. For example, a ReS$_2$ FET with a top gate of the ion-gel electrolyte could provide ultrastrong vertical electric field to actuate the charge transfer.[38]

In the Supporting Information, we also display experiment results of electrically enabled SHG from a five-layer ReS$_2$ flake. Same as the phenomenon observed from the trilayer ReS$_2$, strong SHG signal is obtained from the five-layer ReS$_2$ under the negative gate voltages, which though shows no SHG intrinsically due to the centrosymmetric nature. The $V_G$-dependence of the SHG intensities shows the same functions as that from the trilayer ReS$_2$. When $V_G$ < 0 V, the intensities of SHG decrease to zero first and then increase dramatically. The zero-SHG is obtained at $V_G \approx -30$ V instead at $V_G = 0$ V, which is arisen from the substrate-doping effect. Because of the charge shielding layer formed in the bottom S atom layer under $V_G > 0$ V, the intensities of SHG keep almost unchanged. It confirms the reliable electrical tuning of SHG in odd-number-layered $ReS_2$.

We also carry out the experiments on the tunable SHG in monolayer $ReS_2$, as discussed in the Supporting Information in detail. At $V_G = 0$ V, the monolayer $ReS_2$ intrinsically has no SHG due to its centrosymmetric structure. With $V_G < 0$ V and $V_G > 0$ V, no SHG signal is observed in monolayer $ReS_2$ either. As demonstrated above, the electrically tunable SHG in multilayer $ReS_2$ is determined by the interlayer charge transfer. The monolayer $ReS_2$ has only one molecular layer, there is no interlayer coupling and charge transfer. Due to the strong intralayer covalent bond in the monolayer $ReS_2$, the out-of-plane electric field can not shift the intralayer charge considerably to induce effective second-order nonlinear hyperpolarizability for tunable SHG though the gate voltage is applied.

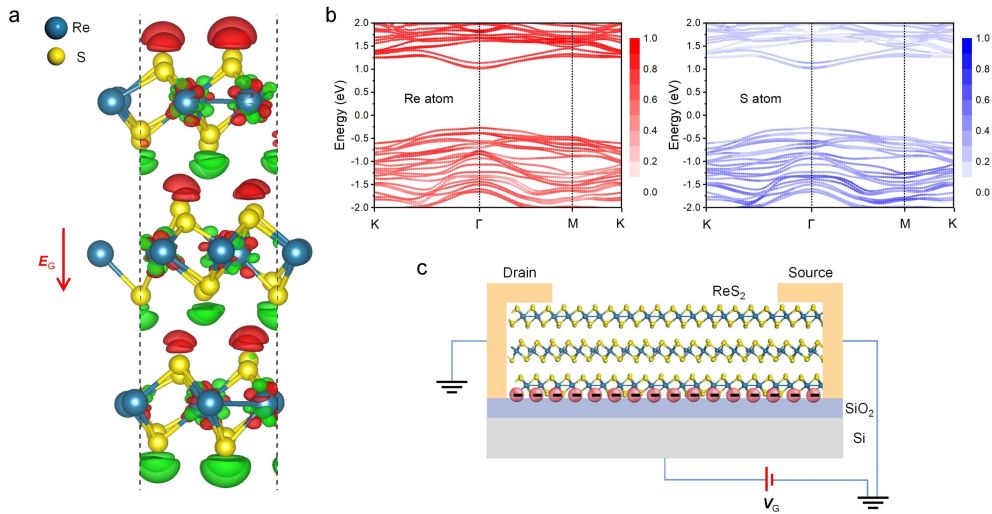

**Figure 3.** (a) Charge density distribution in a trilayer ReS$_2$ under an external electric field pointing downwards, where red (green) region means the gaining (losing) of electrons. (b) Contributions of Re and S atoms on the band structure of a trilayer ReS$_2$. (c) Schematic of the electron accumulation on bottom S atom layer acting as the charge shielding layer that prevents further electric field penetration in the trilayer ReS$_2$ under $V_G$ > 0 V.

The above results indicate that it is possible to control the second-order polarizability in ReS$_2$ flakes with an external electric field, which enables significant SHG from the intrinsically centrosymmetric odd-number-layered ReS$_2$ flakes. Determined by the distorted 1T lattice structure, this electrical tunability of second-order nonlinearity could also be valid on even-number-layered ReS$_2$ flakes, which have intrinsic SHG. As shown in Figure 4, a tetralayer ReS$_2$ FET is fabricated (the inset in Figure 4a) and the $V_G$-dependence of its SHG is characterized carefully. The electrical transfer curve is in Figure 4a, presenting a typical n-type conduction. When $V_G$ = 0 V, a strong intrinsic SHG signal is detected (the blue curve in Figure 4b), which is consistent with our previous work.[21] Upon the application of a negative gate $V_G$ = -60 V, the SHG intensity (the red curve in Figure 4b) is nearly twice as that under $V_G$ = 0 V. The pump power dependences of the SHG signals obtained under $V_G$ = 0 V and $V_G$ = -60 V are examined, as shown in Figure 4c. Both present nearly quadratic dependences. We also measure their laser polarization dependences, as shown in Figure 4d. The same twisted dumbbell shapes with different SHG intensities are observed under $V_G$ = 0 V and $V_G$ = -60 V, indicating the external electric field did not change the crystal lattice structure.

The influence of the external electric field on the SHG of tetralayer ReS$_2$ is further studied, as shown in Figure 4e with the $V_G$-dependent SHG. By applying negative $V_G$, SHG increases gradually. It could be attributed to the strengthened asymmetry of the charge density distribution under the increased negative $V_G$. According to the first-principles calculations, in tetralayer ReS$_2$, there is an intrinsically asymmetric charge distribution induced by the electrons transfer from the bottom layer to the top layer, which is determined by the distorted 1T lattice structure and the layer stacking order. When $V_G$ < 0 V, governed by the external electric field, electrons will shift in the same direction as that at $V_G$ = 0 V. On the whole, compared with the intrinsic tetralayer ReS$_2$, the amount of the net electrons transferred from the bottom layer to the top layer under $V_G$ < 0 V is much greater than that at $V_G$ = 0 V. As a result, the asymmetry of the charge density distribution in the tetralayer ReS$_2$ will be enhanced linearly as the increase of the negative gate voltage. Therefore, the resulted second-order nonlinear hyperpolarizability as well as the SHG increase linearly under the negative gate voltage (see details in the Supporting Information). Note that there is no minimum at $V_G$ = -20 V. Similar to the trilayer ReS$_2$, there should be also a slight n-type doping by the SiO$_2$ substrate in the tetralayer ReS$_2$. However, under the same experimental conditions, the SHG intensity caused by the intrinsic asymmetry in the tetralayer ReS$_2$ is approximately two orders of magnitude greater than that induced by the n-type doping of the SiO$_2$ substrate at $V_G$ = 0 V in trilayer ReS$_2$. Hence, the effect of the slight n-type doping by the SiO$_2$ substrate on SHG intensity at $V_G$ = 0 V in tetralayer ReS$_2$ is very small and can be neglected. Though the negative gate $V_G$ = -20 V would modify the slight n-type doping by the SiO$_2$ substrate, the variation of the SHG is very weak compared with the intrinsically strong SHG. Therefore, in tetralayer ReS$_2$, a minimum SHG at $V_G$ = -20 V was not observed over the strong SHG background. If the $V_G$ is positive, the SHG intensity keeps almost unchanged, which is originated from the charge shielding layer formed by the electrons accumulated on the bottom S atoms. More discussions can be found in the Supporting Information. In addition, the electrically tunable SHG in tetralayer ReS$_2$ could respond to the optical fields in a broad laser wavelength range. As shown in Figure 4f, the electrically tuned SHGs have the similar trend for the pump lasers at the wavelengths between 1500 nm and 1600 nm. Figure 4g displays the SHG spatial mappings of the $ReS_2$ region marked by the red dashed box in the inset of Figure 4a under different gate voltages. Uniform SHG distributions are observed over the whole $ReS_2$ region. The SHG strengths vary with the gate voltages according to the relation shown in Figure 4e.

To show a dynamic modulation behavior of the SHG signal from the tetralayer $ReS_2$, we employ a source-meter to apply a rectangular-wave electrical voltage as the $V_G$ and monitor the corresponding SHG intensities using a photomultiplier tube (PMT). The obtained SHG evolution as a function of gate voltage is shown in Figure 4h. With the periodically varied $V_G$ as 0 V and 60 V, the SHG intensities change subsequently with respect to the external electric field. The rising edge obtained through the experimental result is about $t \approx 0.23$ s, and $f_{3\ dB} \approx 0.44/t \approx 1.91$ Hz. Here, limited by the operation speeds of the employed source-meter and PMT, the maximum SHG modulation speed of the tetralayer $ReS_2$ is not reachable. Since the electrical model for applying electric field on the $ReS_2$ flake is actually an electrical capacitor, the SHG modulation speed is determined by the resistance and capacitance of the device, which is estimated as $f_{3\ dB} \approx 1/2\pi RC$, where $R = 0.56$ MΩ is the resistance of the tetralayer $ReS_2$, $C = \varepsilon_r\varepsilon_0 S/d$ is the gate dielectric capacitance, $\varepsilon_r \sim 3.9$ is the dielectric constant of $SiO_2$, $\varepsilon_0$ is the dielectric constant of free space, $S \approx 3.14$ μm$^2$ is the spot area of the pump laser, and $d = 300$ nm is the thickness of the dielectric layer. Therefore, the 3 dB response bandwidth is $f_{3\ dB} \approx 780$ MHz, which could be greatly strengthened to several GHz by optimizing the structure capacitance.

We also present the electrically tuned SHG in another even-number-layered (8-layer) $ReS_2$ flake, as shown in the Supporting Information. It shows that a similar $V_G$-dependent SHG with that of the tetralayer ReS$_2$, in which the SHG intensities obtained under $V_G$ = 0 V and -40 V have a variation around 3 times.

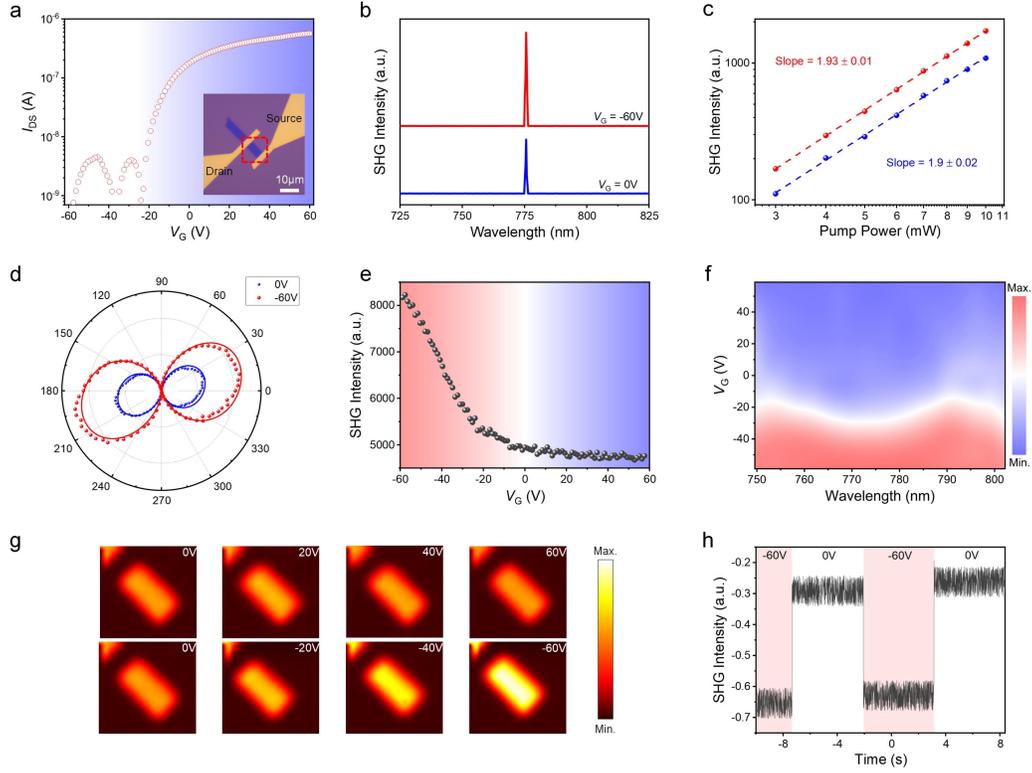

**Figure 4.** (a) Log scale plot of the transfer curve of the tetralayer ReS$_2$ FET. Inset is the optical microscope image of the device. (b) SHG spectra measured from the tetralayer ReS$_2$ FET at $V_G$ = 0 V (blue) and $V_G$ = -60 V (red) under the same experiment conditions. (c) Log-log plot of the pump power dependence of SHG intensities obtained from the tetralayer ReS$_2$ at $V_G$ = 0 V (blue curve) and $V_G$ = -60 V (red curve). (d) Polar plots of parallel components of polarization dependence of the SHG intensities measured from the tetralayer ReS$_2$ at $V_G$ = 0 V (blue curve) and $V_G$ = -60 V (red curve). (e) Gate voltage dependence of the SHG intensities in the tetralayer ReS$_2$. (f) SHG intensities as a function of the gate voltage and the pump wavelength in the tetralayer ReS$_2$. (g) SHG spatial mapping image of the ReS$_2$ region marked by the red dashed

box in the inset of (a) under different gate voltages. (h) Dynamic response of SHG from the tetralayer $ReS_2$ modulated by the gate voltages as 0 V and -60 V, respectively.

The electrically tunable second-order nonlinearity in atomically thin $ReS_2$ flakes is highly promising for the application of nonlinear optoelectronic devices. To illustrate its potential, we further demonstrate an electrically reconfigurable SHG imaging of a $ReS_2$ flake. As shown in Figure 5a, a few-layer $ReS_2$ FET with a split back gate electrodes ($V_{G1}$ and $V_{G2}$) has been fabricated. Two Au metal pads with a gap around 300 nm are first prepared on the $SiO_2$/Si substrate by electron beam lithography and thermal evaporation methods. Then a polymethylmethacrylate (PMMA) film with a thickness of 200 nm is spin-coated on the Au metal pads as the dielectric layer. Finally, a six-layer $ReS_2$ flake and an Au electrode are transferred in sequence onto the PMMA film to complete the fabrication of the dual-gate $ReS_2$ device.

During the measurements of the tuned SHG, we keep the top Au electrode grounded and apply different voltages on the two separated back gates. Figure 5b displays the spatial scanning map of the SHG signal from the device in different gating configurations. With $V_{G1}$ and $V_{G2}$ set as (-60 V, -60 V), (60 V, -60 V), (0 V, -40 V), (-60 V, -40 V), the SHG intensities could be (strong, strong), (weak, strong), (weak, medium), (strong, medium). The regions of the $ReS_2$ flake applied with negative gate voltages have stronger SHG signal, and the strength is proportional to the magnitude of the gate voltage. Without a gate voltage or with a positive gate voltage, the SHG signal is much weaker. These results are consistent with the above explanations. Though measured from the same piece of $ReS_2$ flake, reconfigurable distributions of SHG strengths could be obtained by applying different gate voltages on the two separated gates.

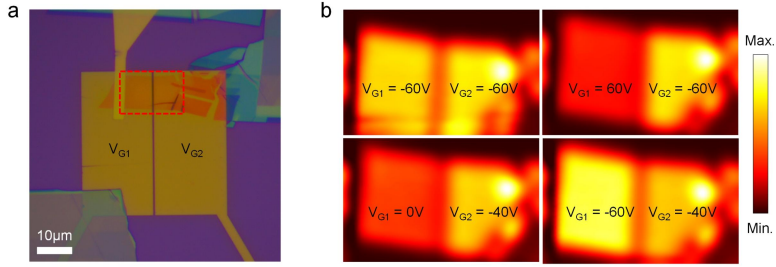

**Figure 5.** (a) Optical microscope image of the ReS$_2$ device with dual back gates. (b) SHG spatial mappings of the ReS$_2$ area marked by the red dashed box in (a), where $V_{G1}$ and $V_{G2}$ are applied with voltages of (-60 V, -60 V), (60 V, -60 V), (0 V, -40 V), (-60 V, -40 V).

**CONCLUSION**

In conclusion, we have demonstrated the electrical tuning of second-order nonlinearity in atomically thin ReS$_2$ flakes. In few-layer ReS$_2$ FETs, out-of-plane electric fields are applied across the ReS$_2$ flakes, which could modify the charge density distributions of each layer via the interlayer charge transfer. Thanks to the distorted 1T crystal structure, the redistributed charges form a modified in-plane second-order hyperpolarizability. The first-principles calculations are carried out to understand the underlying physical mechanism. From odd-number-layered ReS$_2$ flakes with intrinsic centrosymmetric crystal structure, strong SHGs emerge with the actuation of an electric field, which has a strength more than one order of magnitude stronger than that from the monolayer MoS$_2$. In the even-number-layered ReS$_2$ flakes with intrinsic SHG, the external electric field could modulate and enhance the SHGs by several times. From the electrically tuned ReS$_2$ flakes, the dynamic modulation of SHG and reconfigurable SHG imagings were realized successfully. The strong and tunable second-order nonlinearity in atomically thin ReS$_2$ flakes could be exploited to develop electrically controlled nonlinear optoelectronic devices, including nonlinear electro-optic modulators, reconfigurable photodetectors with nonlinear photoresponses, optoelectronic nonlinear logic devices.

**METHODS**

**Device Fabrication.** The atomically thin ReS$_2$ flakes were prepared by mechanically exfoliating a synthetic semiconducting bulk ReS$_2$ material onto a PDMS stamp and then transferred onto a silicon substrate covered with 300 nm thick silicon dioxide (SiO$_2$/Si) by the dry transfer method. The few-layer ReS$_2$ flakes were identified using an optical microscope. The layer numbers were confirmed by the ultra-low frequency Raman spectroscopy and the AFM technique. Au pads with a thickness of 100 nm were defined and deposited on a bare silicon substrate using electron beam lithography and thermal evaporation, which were then mechanically peeled off by PDMS and transferred onto the ends of the ReS$_2$ flakes. The devices were then annealed in the vacuum environment for 2 hours under 150 ºC. Finally, the substrate was stuck onto the metal base by the conductive tape. An ultrasonic aluminum wire welding machine was used to connect the base with the electrodes.

**SHG Measurement.** The measurements of the electrically tuned SHG in few-layer ReS$_2$ were implemented in a home-built vertical microscope setup with the reflection geometry. An OPO pulsed laser with the tunable wavelength ranges between 1400 nm~1800 nm and a repetition rate of 100 MHz was chosen as the fundamental pump radiation. A 50× microscope objective lens with a numerical aperture of 0.75 was employed to focus the pump laser into a spot size about 2 μm on the sample. The SHG signal scattered from the ReS$_2$ sample was collected by the same objective lens. In the signal collection path, a dichroic mirror was used to filter out the pump laser from the SHG signal, which was finally analyzed and detected by a spectrometer (Princeton Instruments, SP 2558 & 100BRX) mounted with a cooled silicon charge-coupled device (CCD) camera. The metal base with the device was kept in a nitrogen cavity during the testing process. The back gate voltage was applied by an electric meter (KEYSIGHT B2912A Precision Source/Measure Unit). The source-drain electrodes were kept grounded during the measurement of electrically tuned SHGs. A home-built labview program realized the synchronization of the electric meter and the spectrometer. The spectrometer collected the SHG signal under the change of the back gate voltage in real time. For the measurements of the electrically tuned SHG under the continuous change of the pump wavelength, the OPO laser wavelength was changed from 1500 nm to 1600 nm with the step about 5 nm and the SHG was collected at each wavelength with the back gate voltage sweeping from -60 V to 60 V.

**The first-principles calculation.** All calculations were performed with the PBE functional[39] and the projected augmented wave potential method[40] in the VASP[41] code based on the density functional theory. The van der Waals interactions were described by optB86b dispersion-corrected scheme.[42,43] The energy cutoff for plane-wave basis was set to 700 eV. The thickness of the vacuum was larger than 20 Å. The atomic structures were relaxed until the forces less than 0.005 eV/ Å. A $k$-point grid of 7×7×1 was used to simplify the first Brillouin zone.

**ASSOCIATED CONTENT**

**Supporting Information**.

Ultra-low wavenumber Raman spectra of trilayer $ReS_2$ and tetralayer $ReS_2$, electrically tunable SHG in five-layer $ReS_2$, electrically tunable SHG in monolayer $ReS_2$, analysis for the electrically tunable SHG in tetralayer $ReS_2$, and electrically tunable SHG in eight-layer $ReS_2$. (Word)

## AUTHOR INFORMATION

**Corresponding Author**

*E-mail: xuetaogan@nwpu.edu.cn

**Author Contributions**

‡Jing Wang and Nannan Han contributed equally to this work

**Notes**

The authors declare no competing financial interest.

## ACKNOWLEDGMENT

The author acknowledges funding from the Key Research and Development Program (2018YFA0307200), the National Natural Science Foundation (61775183, 91950119, 11634010, 61905196, and 62090033), the Key Research and Development Program in Shaanxi Province of China (2020JZ-10), the Fundamental Research Funds for the Central Universities (D5000210905, 310201911cx032 and 3102019JC008). The authors also thank the Analytical & Testing Center of NPU for their assistance in device fabrication and characterizations.

**ToC**

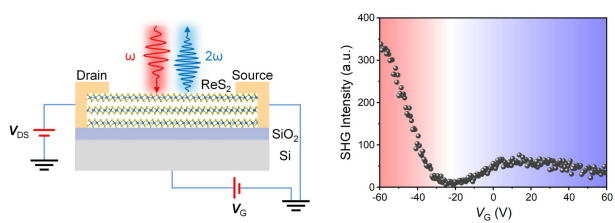